%Paper: gr-qc/9411061
%From: jemal@roxanne.nuclecu.unam.mx (Jemal Guven)
%Date: Wed, 23 Nov 1994 16:37:29 -0600
%Date (revised): Thu, 27 Apr 1995 14:10:07 -0600

%%%%%%%%%%%%%%%%%%%%%%%%%%%%%%%%%%%%%%%%%%%%%%%%%%%%%%%%%%%%%%%%%
%%%%%%%%%%%%%%%%%%%%% Plain TeX file %%%%%%%%%%%%%%%%%%%%%%%%%%%%%%%%
%%%%%%%%%%%%%%%%%%%%%%%%%%%%%%%%%%%%%%%%%%%%%%%%%%%%%%%%%%%%%%%%%

\magnification=1200
\baselineskip=16pt
\hskip 10 cm
CIEA-GR-9409

\hskip 10cm
ICN-UNAM-9406

\vskip .5pc
\centerline{\bf LARGE DEFORMATIONS OF RELATIVISTIC MEMBRANES:}
\vskip.5pc
\centerline{\bf  A GENERALIZATION OF THE RAYCHAUDHURI EQUATIONS}
\vskip1pc
\centerline {\bf Riccardo Capovilla$^{(1)}$ and Jemal Guven$^{(2)}$}
\vskip.5pc
\it
\centerline {$^{(1)}$Departamento de F\'{\i}sica}
\centerline{  Centro de Investigaci\'on y de Estudios Avanzados del
I.P.N. }
\centerline {Apdo. Postal 14-740, 07000 M\'exico, D. F., MEXICO} \rm
\centerline{(capo@fis.cinvestav.mx)}
\vskip.5pc
\it
\centerline {$^{(2)}$ Instituto de Ciencias Nucleares}
\centerline {Universidad Nacional Aut\'onoma de M\'exico}
\centerline {Apdo. Postal 70-543. 04510 M\'exico, D. F., MEXICO} \rm
\centerline{(guven@roxanne.nuclecu.unam.mx)}
%\vfill\eject
\vskip 1pc
\centerline{\bf Abstract}
\vskip.5pc
{\leftskip=1.5cm\rightskip=1.5cm\smallskip\noindent
A coupled system of non-linear partial differential equations
is presented which describes non-perturbatively
the evolution of deformations of a relativistic membrane
of arbitrary dimension, $D$, in an arbitrary background spacetime.
These equations can be considered from a formal point of view
as higher dimensional analogs of the Raychaudhuri equations for
point particles to which they are shown to reduce when $D=1$. For
$D=1$ or $D=2$ (a string), there are no constraints on the initial data.
If $D>2$, however, there will be constraints with a corresponding
complication of the evolution problem. The consistent evolution of the
constraints is guaranteed by an integrability condition which is
satisfied when the equations of motion are satisfied. Explicit
calculations are performed for membranes described by the Nambu action.
\smallskip}
\vfill\eject
\noindent{\bf I INTRODUCTION}
\vskip1pc

A surprising variety of physical systems can be modeled
as relativistic membranes of an appropriate dimension propagating in a
fixed background spacetime. The phenomenological action describing
the dynamics of the membrane is a sum of various relevant scalars
associated with the geometry of its trajectory (worldsheet).
At lowest order this action is proportional to the area of the
worldsheet, the Nambu action [1]. If the approximation stops here, the
classical trajectory of the membrane is an extremal surface of
the background spacetime [2]. At a higher order, one can consider
rigidity corrections quadratic in the extrinsic curvature of the
worldsheet [3].

The dynamics of a Nambu membrane is reasonably well understood.
A large body of information has been accumulated on the dynamics
of geometrically symmetrical extremal configurations [1].
What is less complete is a satisfactory description of
the deformations of relativistic membranes. How does a variation in
the symmetry of the membrane evolve? If the worldsheet develops
a singularity, such as a cusp, how will this singularity evolve? Will
it be smoothed out or will it grow?

It is clear from the outset what criteria  a description
of deformations should satisfy. First, it should be covariant
not only with respect to worldsheet diffeomorphisms, but also
with respect to local rotations of the normals to the
worldsheet itself when the co-dimension of the worldsheet is greater than
one. This is crucial because once we deviate from a
symmetrical configuration the choice of normals will no longer be
obvious. A second requirement is that,
ideally, the description should be independent of the specific
dynamics of the membrane. Apart from obvious economical
considerations, this permits one to isolate
the kinematical features of the deformation, common to all membrane
theories, from those which depend on the
dynamics. Since the dynamics under consideration
is an approximation in the first place, one would like to know
what features of the evolution of deformations are
influenced by a change in the dynamics.

Various aspects of this problem have been addressed recently by several
authors. Garriga and Vilenkin described the evolution of small disturbances
propagating on planar and spherically symmetrical Nambu membranes in
background Minkowski and de Sitter spacetimes [4]. Following this work,
Guven [5] and almost simultaneously, Carter as well as Frolov
and Larson [6] approached the problem of small deformations of a
Nambu membrane in a manifestly covariant way, independent of the
particular symmetry of the defect, and of the background spacetime.
In [5], the role of the twist potential in ensuring
manifest covariance under normal rotations was made explicit. The
deformation is described by a set of massive scalar fields,
which satisfy a coupled system of linear wave equations.
The scalar fields are the projection of the infinitesimal
deformation in the embedding function describing the
worldsheet onto each normal direction.
The effective mass matrix is the sum of a term
quadratic in the extrinsic curvature, and a matrix
of curvature projections. This framework was subsequently generalized
by the authors to permit the stability analysis of any
membrane described by a local action constructed from any worldsheet
scalars [7]. This was done by shifting the focus from the particular
dynamics of an extremal surface to a systematic kinematical
description of the deformation of the worldsheet intrinsic and
extrinsic geometries.

The progress we have described, has been made entirely
in the examination of small deformations.
If the membrane is described by the Nambu action there is no
energy penalty prohibiting the formation of kinks or cusps ---
we should therefore not be surprised to find that structures
like these form in the course of the evolution. Unfortunately, the analysis
of such structures lies outside the scope of perturbation theory.
A formalism permitting us to examine large deformations is required.

{}From a formal point of view, the equations describing the evolution of
small deformations of the worldsheet are higher dimensional analogs of the
Jacobi equations describing the infinitesimal separation of
neighboring timelike geodesics [8]. These higher dimensional analogs
possess, however, an interpretation without any one-dimensional
analog in that they describe a physical stretching of the
membrane itself.

In the prototype of a geodesic curve, an alternative non-perturbative
framework for describing the focusing of trajectories is provided by
the Raychaudhuri equations [8,9].  These are a system of coupled
non-linear ordinary differential equations describing the evolution
along a curve of the orthogonal deformations of its velocity
vector.  By examining the trace of these equations
over the normal directions, which describes the expansion
in the volume occupied by a given pencil of geodesics, it is possible
to show that under very reasonable assumptions on the material
sources, spacetime will generally be geodesically incomplete. As such,
the Raychaudhuri equations constitute one of the cornerstones of
the classical singularity theorems in general relativity.

If the Raychaudhuri equations are any guideline, we should shift our
focus from the deformation in the embedding function describing
the worldsheet of the membrane to the deformation in the tangent
vectors to this surface [10]. In this way we derive an analogous
non-perturbative, and non-linear, coupled system of partial differential
equations to describe the deformation of higher dimensional surfaces.
Unlike the Jacobi equation, where the formal generalization is in
some sense the obvious one, the generalization of the
Raychaudhuri equations is extremely non-trivial.

The particular value of this set of equations is that they
provide us with an analytically tractable procedure for
examining various peculiarities of the dynamics of
relativistic membranes whose description is beyond the scope of
perturbation theory. Physical applications will be considered elsewhere
[11].

The content of this article is as follows:

To establish our notation, we begin in Sect. II by summarizing
the classical kinematical description of an embedded
timelike surface (worldsheet) of dimension $D$, in a fixed
background spacetime of dimension $N$, in terms of its
intrinsic and extrinsic geometry. The latter is characterized
completely by the extrinsic curvature and the extrinsic
twist potential. This worldsheet will be generated from some initial
configuration of the membrane appropriate to the truncation
of the action describing its dynamics.

We now want to consider the evolution of a deformation of this worldsheet.
We begin by providing a purely kinematical description of the deformation
of the worldsheet, a non-perturbative analog of
the analysis performed in Ref. [7]. In this way we identify the
structures that characterize the deformation. When $D=1$,
these are the worldsheet scalars $J^{ij}$ constructed by
taking the projection onto the $j^{\rm th}$ normal
of the gradient of the tangent vector to the curve along the
$i^{\rm th}$ normal. When $D>1$, there will be one object of this
kind for each tangent vector to the surface, $J_a{}^{ij}$
($a,b,\cdots=0,1,\cdots, D-1$). In addition to such
straightforward generalizations from one to higher dimensions
there will be a new structure without any one-dimensional
analog, which we interpret geometrically as the
deformation of the worldsheet connection preserving manifest
covariance under tangent frame rotations after deformation.

In Sect. III, a `naive' generalization of the Raychaudhuri equations
which reduces to the familiar prototype in  one-dimension is provided.
In Sect. IV we point out the inadequacies of this simpleminded
generalization. If our system of equations is to possess any predictive
power, it is important that it possess a Cauchy formulation,
modulo the membrane dynamics, so that the evolution of independent
initial deformations can be tracked. We encounter various obstacles
to the implementation of such a formulation.

Firstly, the source for the deformation in the equation of motion
must not involve unknown elements. To obtain a consistent system of
equations, we need then  to form suitable linear combinations of the
naive generalization of the Raychaudhuri equations, such that these
elements are eliminated from
the source. One set of linear combinations is an antisymmetric
sum with respect to worldsheet indices. This linear combination eliminates
these unknowns no matter what the background dynamics is. In Sect. V,
we consider extremal surfaces, and for this case the remaining
linear combination is a trace.

We are still not out of the woods, however. This is because when
$D>2$ not all of our equations are dynamical. The non-dynamical
equations must be considered as constraints on the initial
data, {\it i.e.} the deformations on some spacelike hypersurface
of the worldsheet. These data are not freely specifiable. The existence of
constraints complicates the implementation of a
Cauchy solution, for we need to ensure that they be preserved by the
evolution. This could require us to impose a non-trivial integrability
condition on the solution, thereby further complicating the solution
of the Cauchy problem. We demonstrate, however, that the integrability
conditions we require are trivially satisfied, modulo the
differential Bianchi identities on the curvature associated with the
twist potential introduced in Sect. II.
Once the constraints are satisfied by the initial data,
the equations of motion will ensure that they continue to be satisfied at
all subsequent times. We point out three important dimensional exceptions:
the cases of a point particle, $D=1$, and of a string, $D=2$,
where there are no constraints on the initial data, and that of a
hypersurface, $D=N-1$, where the constraint reduces to a
condition on the rotationality of the spatial initial data.

Now that we have a consistent generalization of the Raychaudhuri
equations to higher dimensions, we focus on the generalized expansion,
given by tracing $J_{a\,ij}$ over normal indices, {\it i.e.} $
J_a{}^i{}_i \equiv \Theta_a$. The anti-symmetric linear equation implies
that $\Theta_a = \partial_a \Upsilon$. For extremal surfaces, inserting
this in the traced evolution equation for $\Theta_a$ gives
Eq. (5.15) in the text,

$$
\Delta \Upsilon + {1 \over N-D} \partial_a \Upsilon \partial^a \Upsilon
- \Lambda^2 + \Sigma^2 - M^2 = 0\,.
$$
where $\Lambda^2 , \Sigma^2, M^2$ are worldsheet scalars defined
in the text.
This equation generalizes the familiar first order ordinary differential
equation for the expansion of neighbouring geodesics.

In Sect. VI, we compare our results to perturbation theory.
For an extremal surface, the `naive' truncation consisting only of the
traced equations (or the appropriate set of equations if the dynamics is
not extremal) permits us to recover the coupled linear scalar equations
of motion describing the evolution of an infinitesimal
deformation of the extremal worldsheet derived in Refs.[5,7].
The anti-symmetric Raychaudhuri equations play no role in perturbation
theory.

Finally, we conclude in Sect. VII, with a  discussion of our results.

For simplicity, we confine our attention to closed surfaces
without physical boundaries. All considerations are local.

%\vfill\eject
\vskip2pc
\noindent{\bf II MATHEMATICAL PRELIMINARIES}
\vskip1pc

\noindent{\bf II.1  Geometry of the Worldsheet}
\vskip1pc

We consider an oriented worldsheet $m$ of dimension $D$
described by the embedding,

$$
x^\mu=X^\mu(\xi^a)\,,\eqno(2.1)$$
($\mu, \nu, \cdots = 0, 1, \cdots, N-1$, and $a, b, \cdots = 0, 1, \cdots,
D-1$)
in a  spacetime $\{M, g_{\mu\nu}\}$ of dimension $N$.
We adopt the signature $\{-,+, \cdots ,+\}$ for $g_{\mu\nu}$.
The $D$ vectors

$$
e_a =X^\mu_{,a}\; \partial_\mu\eqno(2.2)$$
form a basis of tangent vectors to $m$ at each point of $m$.
The metric induced on the worldsheet is then
given by,

$$
\gamma_{ab}= X^\mu_{,a} X^\nu_{,b}\,g_{\mu\nu} =
g(e_a,e_b)\,.\eqno(2.3)$$

To facilitate comparison with the one dimensional Raychaudhuri
prototype, where the single tangent vector is the unit velocity
vector along the particle worldline, it is useful, if not essential,
to replace the coordinate tangent basis, $\{ e_a \}$,
by an orthonormal basis of tangent vectors,
$\{ E_a \}$,

$$
g(E_a, E_b)=\eta_{ab}\,. \eqno(2.4)$$
In order to avoid cluttering our equations,
we continue to use latin letters for orthonormal indices.

We assume that the worldsheet is timelike everywhere, and that
we can always consistently choose one timelike tangent vector field, $E_0$.
That the vectors $\{ E_a \}$
form a surface is encoded in an integrability
condition, the closure of their commutator algebra, $[E_a, E_b]$, by
Frobenius theorem [2].

The $i^{\rm th}$ unit normal to the worldsheet,
($i, j, \cdots = 1, 2, \cdots, N-D$) is denoted with $n^i$, and is defined up
to a local
$O(N-D)$ rotation by

$$
g(n^i, n^j ) = \delta^{ij}\,, \;\;\;\;\;\; g(n^i, E_a) = 0\,.
\eqno(2.5) $$
Normal vielbein indices are raised and
lowered with $\delta^{ij}$ and $\delta_{ij}$, respectively, whereas
tangential indices are raised and lowered with $\eta^{ab}$ and
$\eta_{ab}$, respectively.

We define the worldsheet projections of the
spacetime covariant derivatives, $D_a= E^\mu{}_a D_\mu$,
where $D_\mu$ denotes the (torsionless) covariant derivative
compatible with the spacetime metric $g_{\mu \nu}$.
Let us  consider the
worldsheet gradients of the basis vectors
$\{E_a, n^i\}$.
Since they are spacetime vectors, they  can always
be decomposed with respect to the orthonormal basis $\{ E_a , n^i \}$
[2], as,

$$
\eqalignno{D_a E_b =& \gamma_{ab}{}^c E_c  - K_{ab}{}^i n_i\,, &(2.6.a) \cr
	   D_a n^i =& K_{ab}{}^i E^b + \omega_a{}^{ij} n_j\,. &(2.6.b)\cr}$$
These kinematical expressions, which describe the extrinsic
geometry of the worldsheet, are generalizations of the classical
Gauss-Weingarten equations. The $\gamma_{ab}{}^c$ are the
worldsheet Ricci rotation coefficients,

$$
\gamma_{abc} \equiv g(D_a E_b,E_c)=-\gamma_{acb}\,.\eqno(2.7)$$
The quantity $K_{ab}{}^i$ is the $i^{\rm th}$
extrinsic curvature of the worldsheet,

$$
K_{ab}{}^{i} \equiv - g(D_a E_b,n^i) =K_{ba}{}^i\,.\eqno(2.8)$$
Its symmetry in the tangential indices is
a consequence of the integrability of the $\{E_a\}$.

The twist potential
of the worldsheet is defined by
$$\omega_a{}^{ij} \equiv g(D_a n^i,n^j) =-\omega_a{}^{ji}\,.\eqno(2.9)$$
With respect to a normal   frame rotation,
$n^i\to O^{i}{}_{j} n^j$,  $\omega_a{}^{ij}$
transforms as a connection,
$\omega_a\to O \omega_a O^{-1} + O_{,a} O^{-1}$.
As discussed {\it e.g.} in Ref.[5], it is therefore associated with the
gauging of normal frame rotations. It is desirable to implement normal
frame covariance in a manifest way. We therefore introduce
a worldsheet covariant derivative, defined on fields
transforming as tensors under normal frame rotations as follows,

$$
\tilde\nabla_a \Phi^{i_1\cdots i_n} =
\nabla_a \Phi^{i_1\cdots i_n} - \omega_a{}^{i_1 j} \Phi_j{}^{i_2\cdots i_n}
-\cdots- \omega_a{}^{i_n j} \Phi^{i_1\cdots i_{n-1}}{}_j\,,\eqno(2.10)$$
where   $\nabla_a$ is the intrinsic worldsheet covariant derivative.

The embedding $X^\mu(\xi)$ is overspecified by Eqs. (2.6). There are
integrability conditions: the intrinsic and extrinsic geometry
must satisfy the Gauss-Codazzi, Codazzi-Mainardi, and Ricci
integrability conditions, given, respectively, by,

$$
\eqalignno{g( R(E_b,E_a)E_c,E_d) &= {\cal R}_{abcd} -
K_{ac}{}^{i} K_{bd\,i} + K_{ad}{}^{i} K_{bc\, i}\,, &(2.11a) \cr
g( R(E_b,E_a)E_c, n^{i}) &=  \tilde\nabla_a K_{bc}{}^{i} -
\tilde\nabla_b K_{ac}{}^{i}\,, &(2.11b) \cr
g( R(E_b,E_a)n^{i},n^{j}) &=
\Omega_{ab}{}^{ij} - K_{ac}{}^{i} K_b{}^{c\,j}
+ K_{bc}{}^{i} K_a{}^{c\,j}\,. &(2.11c)\cr}$$
We use the notation
$g(R(Y_1, Y_2)Y_3,Y_4) =
 R_{\alpha\beta\mu\nu}Y_1^\mu Y_2^\nu Y_3^\beta Y_4^\alpha$.
$R^\alpha{}_{\beta\mu\nu}$ is the Riemann tensor of the
spacetime covariant derivative $D_\mu$, whereas ${\cal R}^a{}_{bcd}$ is
the Riemann tensor of the worldsheet covariant derivative $\nabla_a$.
$\Omega_{ab}{}^{ij}$ is the curvature associated with the twist
potential $\omega_a{}^{ij}$, defined by,

$$
\Omega_{ab}{}^{ij}=\nabla_b \omega_a{}^{ij} - \nabla_a \omega_b{}^{ij}
- \omega_b{}^{ik} \omega_{a\,k}{}^{j} + \omega_a{}^{ik}
\omega_{b\,k}{}^{j}\,.\eqno(2.12)$$
Since it is a curvature, it satisfies the differential Bianchi identity

$$
\tilde\nabla_{[c} \Omega_{ab]}{}^{ij} = 0\,. \eqno(2.13)$$

The gauge invariant measure of the twist is completely determined by the
extrinsic curvature, and the spacetime Riemann tensor.
The twist potential can be gauged away locally if and only if
$\Omega_{ab}{}^{ij}=0$. Setting $\Omega_{ab}{}^{ij}=0$ in Eq.(2.11c),
therefore, provides the necessary and sufficient condition that
the twist potential can be gauged away, in terms of an
equality between a  projection of the spacetime
Riemann tensor, and an anti-symmetric sum of the square of the
extrinsic curvature.

\vskip1pc
\noindent{\bf II.2  Geometry of a Deformed Worldsheet}
\vskip1pc

The Gauss-Weingarten Eqs.(2.6) describe a single surface embedded
in spacetime. Let us  consider now a one-parameter family of neighboring
surfaces, $x^\mu=X^\mu(\xi^a,s)$. To provide a measure of the relative
displacement of such surfaces, we consider the gradients of the spacetime
basis $\{E_a,n^i\}$ along the directions orthogonal to the worldsheet. Let
$\delta = \partial_s$. We define $\tilde\delta^\mu =(g^{\mu\nu}
-E^\mu{}_a E^{\nu\,a})\delta_\nu$, and
$D_{\tilde\delta} =\tilde\delta^\mu D_\mu$. Now,
the two measures of the orthogonal deformation of $m$ are
$g(n_i,D_{\tilde\delta} E_a)$, and $g(n_i, D_{\tilde\delta} n_j)$.
In particular, suppose that $\tilde\delta= n_i$. Then,
with $D_i =n^\mu{}_i D_\mu$,
the gradients of the spacetime basis $\{E_a,n^i\}$ along the
directions orthogonal to the worldsheet, can be expressed as,

$$
\eqalignno{D_i E_a &= S_{ab\,i} E^b + J_{a\,ij} n^j\,, &(2.14a)\cr
  D_i n_j &= - J_{a\, ij} E^a    + \gamma_{ij}{}^k n_k\,, &(2.14b)\cr}$$
These are the analogues of the Gauss-Weingarten equations, Eqs. (2.6),
along the distribution spanned by the $\{n^i\}$.
Note that one can obtain integrability conditions analogous
to Eqs. (2.11), but we will not need them here.

The quantity $J_{a}{}^{ij}$, which  plays a central role in the
description of the deformations of the worldsheet, is defined by,

$$
J_a{}^{ij} \equiv g(D^i E_a,n^j)
\,.\eqno(2.15)$$
In general, it does not possess any symmetry under interchange of
the normal indices $i$ and $j$, reflecting the fact that
the $\{n^i\}$ (unlike the $\{E_a\}$) do not generally form
an integrable distribution. It is the analogue of
$K_{ab}{}^i$ in the Gauss-Weingarten equations.

The quantity $S_{ab}{}^i$ is defined by

$$
S_{ab}{}^i \equiv g(D^i E_a,E_b)= - S_{ba}{}^{i}\,.\eqno(2.16)$$
For an embedded curve, $S_{ab}{}^i=0$, identically.
It is the analogue of the extrinsic twist potential,
$\omega_a{}^{ij}$, in Eqs. (2.6). Note that
under a tangent frame orthonormal
rotation, $E^a \to O^a{}_b E^b$, $S_{ab}{}^i$ transforms
as a connection,
$S_{i}\to O S_i O^{-1} + O_{,i} O^{-1}$.
$S_{ab}{}^{i}$ is the deformation of the worldsheet connection
associated with the gauging of local tangent frame rotations.  We
can introduce an associated covariant derivative $\tilde\nabla_i$
in a manner directly analogous to Eq.(2.10) by

$$
\tilde\nabla_i\Phi_{a_1\cdots a_n} =
\nabla_i\Phi_{a_1\cdots a_n} -
S_{a_1 b\,i}\Phi^b{}_{a_2\cdots a_n}- \cdots -
S_{a_n b\,i}\Phi_{a_1\cdots a_{n-1}}{}^b\,,\eqno(2.17)$$
where

$$
\nabla_i \Phi_j =D_i\Phi_j -\gamma_{ij k}\Phi^k\,,\eqno(2.18)$$
is the normal covariant derivative, and,

$$
\gamma_{ijk} \equiv g(D_i n_j,n_k)=-\gamma_{ikj}\,,\eqno(2.19)$$
are the Ricci rotation coefficients associated with the
normal basis.
%The two measures of the orthogonal separation of neighbouring surfaces
%are the quantities $J_{a}{}^{ij}$, and $S_{ab}{}^i$.

\vskip2pc
\noindent{\bf III NATURAL GENERALIZATION OF THE RAYCHAUDHURI EQUATIONS}
\vskip1pc

For a geodesic curve, the Raychaudhuri equations describe the
evolution of $J^{ij}\equiv J_{0}{}^{ij}$, connecting neighboring geodesics
along the curve, given specified values for $J^{ij}$ at some
initial instant. These are ordinary differential equations. Their
generalization to higher dimensional surfaces will clearly involve
partial differential equations.

The natural generalization of the proper time derivative
along the trajectory, $d_s J_{ij}$, in the one dimensional context, is
given by the covariant worldsheet derivatives $\tilde\nabla_b J_{a}{}^{ij}$.
Therefore, to generalize the Raychaudhuri equations,
we evaluate this quantity. We find,

$$
\tilde \nabla_b J_{a}{}^{ij} =
- \tilde \nabla^i K_{ab}{}^{j}
- J_{b}{}^i{}_k J_{a}{}^{kj}- K_{bc}{}^i K_a{}^{c\,j} +
  g (R ( E_b , n^i ) E_a , n^j )
\,.\eqno(3.1)$$
The details of the evaluation of $\tilde\nabla_b J_{a}{}^{ij}$
are contained in Appendix A. We emphasize that the evaluation does
not depend on the membrane equations of motion.

When $D=1$, we reproduce the Raychaudhuri equations for a curve.
For then there is only one tangent vector, the unit timelike
vector $E_0=V$. Now $K_{00}{}^i= - K^i= g(n^i, D_V V)$. In addition,
$S_{00}{}^i =0$ by anti-symmetry, and we can always
orient the normals along the curve such that
the extrinsic twist potential vanishes, $\omega_0{}^{ij}=0$ [5].
(In Ref. [8], this is accomplished by Fermi-Walker transporting the
normals.) With respect to proper time $s$ along a physical trajectory,
Eq.(3.1) assumes the form,

$$
{d J^{ij}\over ds} + \nabla^i K^j
+ J^{i}{}_k J^{kj} + K^i K^j = g ( R (V, n^i) V, n^j)
\,,\eqno(3.2)$$
which agrees, before symmetrization, with Eq.(4.25)  in Ref.[8].

To establish contact with the literature on the Raychaudhuri equations,
we note that the analogs of the quantities which more frequently
appear in the literature, are the spacetime tensors

$$
J_{\mu\nu\,a}= n_\mu{}^i n_\nu{}^j J_{a\,ij}
= H_{\mu}{}^\alpha H_{\nu}{}^\beta D_\alpha E_{\beta\,a}\,,\eqno(3.3)$$
where,

$$
H_{\mu\nu}= n^i_{\mu} n_{i\,\nu} = g_{\mu\nu} -E_\mu{}^a E_{\nu\,a}
\eqno(3.4)$$
is the projection orthogonal to the worldsheet.

\vskip2pc
\noindent{\bf IV DIFFICULTIES WITH THE GENERALIZATION }
\vskip1pc

Given some initial conditions on an initial spacelike hypersurface,
the membrane equations of motion, together with the integrability
conditions, Eqs. (2.11), determine uniquely  the embedded worldsheet
described by Eq.(2.1). This will in turn determine both the
intrinsic geometry $\gamma_{ab}$, or $\{ E_a \}$,
and the extrinsic geometry, as characterized by $K_{ab}{}^i$
and $\omega_a{}^{ij}$.

Suppose we also prescribe some initial values for the $J_a{}^{ij}$ on this
spacelike hypersurface. Is it then possible to determine $J_a{}^{ij}$
at subsequent times using Eq.(3.1)?

Unfortunately, there are at least two obstructions to the
implementation of a Cauchy solution of Eq.(3.1).

The first problem is one which must be confronted even when
$D=1$. The term $\tilde\nabla^i K_{ab}{}^j $, which is not determined on
the worldsheet by the equations of motion, appears as a source for
the worldsheet derivative of $J_a{}^{ij}$ in  Eq.(3.1).
In the traditional one-dimensional application, this does
not present any problem  because one is concerned there
only with geodesics satisfying $K^i=0$. Therefore this term
vanishes, and along with it the dependence on
the normal frame Ricci rotation coefficients, $\gamma_{ij}{}^k$.
If, however, the curve satisfies some other equation of motion
the source is just as much an unknown as is $J^{ij}$, and the
equations useless in practice as they stand.
In general, when $D>1$, we will not only have the normal gradients of
$K_{ab}{}^i$ to contend with in the source term, but also
the connection $S_{ab}{}^i$.

We need therefore to find appropriate linear combinations of
Eq.(3.1), such that the troublesome source term is eliminated,
modulo the equations of motion of the membrane. This is
unfortunate, since, our ideal is a description
of deformations of the worldsheet which is independent of the dynamics,
in a way analogous to the perturbative analysis of Ref. [7]. The
presence of the source term does not leave us any choice,
but to exploit the equations of motion to eliminate it.
We focus on the case of a membrane satisfying the Nambu dynamics,
{\it i.e.} extremal membranes. The generalization to membranes
described by more complicated dynamical rules will be briefly
sketched in the conclusions.

\vskip 2pc
\noindent {\bf V RAYCHAUDHURI EQUATIONS FOR EXTREMAL MEMBRANES}
\vskip 1pc

In this section, we consider membranes which satisfy the
Nambu dynamics. The Nambu action is proportional to the
worldsheet area,

$$
S [X^\mu , X^\mu{}_{,a} ] = - \sigma \int_m d^D \xi \sqrt{- \gamma}
\,, \eqno(5.1)$$
where the constant $\sigma$ is the membrane tension.
The equations of motion are given by [2],

$$
K^i \equiv \gamma^{ab} K_{ab}{}^i = 0
\,.\eqno(5.2)$$

We want to eliminate the term in Eqs. (3.1) involving normal derivatives
of $K_{ab}{}^i$, using Eqs. (5.2). The obvious thing to do is to
consider the linear combination of Eqs. (3.1) obtained by tracing
with $\gamma^{ab}$ over worldsheet indices,

$$
\tilde \nabla_a J^{a\,ij}
+ J_{a}{}^i{}_k J^{a\,kj} + K_{ac}{}^i K^{ac\,j} =
  g (R ( E_a , n^i ) E^a , n^j )
\,,\eqno(5.3)$$
where we have used Eq. (5.2) to eliminate the source term, which
now takes the form $\tilde\nabla^i K^j $.

When $D=1$, this equation reduces to the familiar Raychaudhuri equations
Eq. (3.2). When $D>1$, Eqs.(5.3) assume the form of
a system of coupled continuity-type equations involving the
time derivative of $J_0{}^{ij}$ (a gradient along the
timelike tangent vector, $E_0$), and worldsheet spatial
gradients of the remaining $J_{a}{}^{ij}$.

Now, to evolve initial data, we need one equation to evolve each
variable $J_{a}{}^{ij}$. The traced equation (5.3) does this for
$J_0{}^{ij}$. To complete the description of the evolution,
we need corresponding evolution equations for the
remaining $J_{A}{}^{ij}$, where capital Latin letters denote
worldsheet {\it spatial} indices, $(A,B,\cdots = 1,2,\cdots,D-1 )$.

An important observation towards this goal is the recognition that
the source term in Eqs. (3.1) is symmetric in the worldsheet
indices $a$ and $b$. If we now anti-symmetrize Eq.(3.1) with respect to
its worldsheet indices, we get,

$$
\tilde \nabla_a J_{b}{}^{ij} -\tilde \nabla_b J_{a}{}^{ij} = G_{ab}{}^{ij}
\,,\eqno(5.4)$$
where we have defined,

$$
G_{ab}{}^{ij} \equiv
- J_{a}{}^i{}_k J_{b}{}^{kj}- K_{ac}{}^i K_b{}^{c\,j} +
  g (R ( E_a , n^i ) E_b , n^j ) - ( a \leftrightarrow b )
\,.\eqno(5.5)$$
The source term has been cancelled out independent of the
background dynamics. We note that when $D=1$, Eqs.(5.4) are vacuous.
When $D=N-1$, for a hypersurface, $G_{ab}{}^{ij}$ and the extrinsic
twist potential $\omega_a{}^{ij}$ vanish identically, and Eq.(5.4) reads,

$$
\partial_a J_b - \partial_b J_a = 0
\,,\eqno(5.4^\prime)$$
with $J_a \equiv J_a{}^{11}$.  $J_a$ is rotationless.

The right hand side of Eq. (5.4) can be put in a simpler form. Using
the Ricci integrability conditions Eq. (2.11c), the anti-symmetric
sum of quadratics  in the extrinsic curvature can be
identified with $\Omega_{ab}{}^{ij}$, modulo a projection of the
space-time Riemann tensor. This gives

$$
G_{ab}{}^{ij} = - J_{a}{}^i{}_k J_{b}{}^{kj} + J_{b}{}^i{}_k J_{a}{}^{kj}
- \Omega_{ab}{}^{ij}
\,, \eqno(5.6)$$
where the spacetime Riemann tensor projections vanish, because of
the spacetime  cyclic Bianchi identities, $R_{\mu [\nu \rho \sigma]} = 0$.

At this point, to facilitate our counting of the evolution equations,
let us separate the timelike component of Eq. (5.3), from
the spatial components,

$$
\eqalignno{ \tilde \nabla_0 J_{B}{}^{ij} - \tilde \nabla_B J_{0}{}^{ij}
&= G_{0B}{}^{ij}\,, &(5.7a) \cr
\tilde \nabla_B J_{A}{}^{ij} -\tilde \nabla_B J_{A}{}^{ij} &=
G_{BA}{}^{ij}\,. &(5.7b) \cr } $$
For each pair $(ij)$, Eq. $(5.7a)$ give the required $D-1$ evolution
equations for the $J_A{}^{ij}$, With the trace equations (5.3), they
provide the desired number of equations of evolution.

When $D>1$, for each pair $(ij)$, Eq. $(5.7.b)$ are $(D-1)(D-2)/2$ equations,
that involve only spatial gradients, and, as such, do not evolve
initial data. If these equations are to be interpreted as
constraints on the initial data, it is essential that
they be preserved by the evolution. In the worst possible
scenario, this could require us to impose integrability
conditions on the solution, somewhat analogous to the second class
constraints in Dirac's classification of constrained
dynamical systems [12], thereby further complicating the
solution of the Cauchy problem. What is remarkable is that
the integrability conditions are, in fact, trivially satisfied.
Let us examine how this happens.

Formally, Eqs.(5.3) and (5.4)
represent an over-determined system. Ignoring the
initial value interpretation of this system for the moment,
a solution will exist if and only if  the following
integrability conditions are satisfied,

$$
2\tilde\nabla_{[a}\tilde\nabla_b J_{c]}{}^{ij}
=\tilde\nabla_{[a} G_{bc]}{}^{ij}\,.\eqno(5.8)$$
When $D = 2$, {\it i.e.} for a string, these integrability conditions
are vacuus, and Eqs.(5.3) and (5.4) therefore form a consistent
system of partial differential equations for $J_{a}{}^{ij}$.

When $D > 2$,  these integrability conditions are identically satisfied.
To see this, note that the left hand side gives,

$$
2 \tilde\nabla_{[a}\tilde\nabla_b J_{c]}{}^{ij} =
 \Omega_{[ab}{}^{ik} J_{c]\,k}{}^j +  \Omega_{[ab}{}^{jk} J_{c]}{}^i{}_k
\,,$$
and that the right hand side gives the same,

$$
\eqalign {\tilde\nabla_{[a} G_{bc]}{}^{ij}
&= - \tilde\nabla_{[a} \Omega_{bc]}{}^{ij}
-2(\tilde\nabla_{[a} J_b{}^{ik}) J_{c]\,k}{}^j
- 2 J_{[b}{}^{ik} ( \tilde\nabla_a  J_{c]\,k}{}^j ) \cr
&= - G_{[ab}{}^{ik} J_{c]\,k}{}^j - J_{[b}{}^{ik} G_{ac]\,k}{}^j \cr
&= + \Omega_{[ab}{}^{ik} J_{c]\,k}{}^j + J_{[b}{}^{ik} \Omega_{ac]\,k}{}^j
\cr
&= + \Omega_{[ab}{}^{ik} J_{c]\,k}{}^j +  \Omega_{[ab}{}^{jk}
J_{c]}{}^i{}_k\,. \cr }
$$
We have used the Bianchi differential identity Eq. (2.12) in the first line,
and in the third line a combination cubic in $J_a{}^{ij}$
vanishes identically.

The integrability conditions that might have obstructed the
implementation of the Cauchy problem are then trivial. It is therefore
true that if the constraints are satisfied on an initial
spacelike hypersurface, and the equations of evolution are
satisfied, then the constraints will continue to be satisfied at all
subsequent times. In Appendix B,  we illustrate this, using a
simple example in which the intrinsic geometry of the worldsheet is
flat, and the extrinsic twist potential vanishes.

To summarize, for an extremal membrane, the generalization of the
Raychaudhuri equations for a curve is given by Eqs. (5.3) and Eqs.
(5.4). These equations decribe the evolution of the deformation
of the worldsheet, $J_a{}^{ij}$.

The quantity $J_a{}^{ij}$ is difficult to work with.
By analogy with continuum mechanics, we can decompose
$J_{a}{}^{ij}$ into its symmetric and antisymmetric parts
with respect to the normal indices, $\Theta_{a}{}^{ij}$
and $\Lambda_{a}{}^{ij}$, respectively,

$$
J_a{}^{ij} = \Theta_a{}^{ij} + \Lambda_{a}{}^{ij}
\,.\eqno(5.9)$$
We further decompose $\Theta_{a}{}^{ij}$ into its tracefree and
trace parts,

$$
\Theta_{a}{}^{ij} = \Sigma_{a}{}^{ij} + {1\over N-D}
\delta^{ij} \Theta_a
\,.\eqno(5.10)$$
In one dimension $\Theta$, $\Sigma^{ij}$ and $\Lambda^{ij}$
describe respectively the expansion, the shear, and the
vorticity of a trajectory with respect to neighbouring trajectories.
No such clear interpretation appears to be available in higher dimensions.

We consider now the trace, tracefree, and antisymmetric parts
of the Raychaudhuri equations for an extremal membrane, Eqs. (5.3) and (5.4),
in order to obtain `equations of motion' for
$\Theta_a$, $\Theta_{a}{}^{ij}$, and $\Lambda_{a}{}^{ij}$.

{}From the traced Raychaudhuri equation, Eq. (5.3), one finds

$$
\eqalignno
{\tilde\nabla_a \Lambda^{a\,ij} &+ \Lambda^{a\,k[i} \Lambda_a{}^{j]}{}_k
 + \Sigma^{a\,k[i} \Sigma_a{}^{j]}{}_k
- 2 \Lambda^{a\,k[i} \Sigma_a{}^{j]}{}_k
+ {2 \over N-D} \Lambda_a{}^{ij} \Theta^a = 0\,,  &(5.11.a) \cr
\nabla_a \Theta^a &- \Lambda^{a\,ij} \Lambda_{a\,ij}
+ \Sigma^{a\,ij} \Sigma_{a\,ij} + {1 \over N-D} \Theta_a \Theta^a
- (M^2)^i{}_i = 0\,, &(5.11.b) \cr
\tilde\nabla_a \Sigma^{a\, ij} &+ (\Lambda^{a\,ik} \Lambda_{a\,k}{}^j
+ \Sigma^{a\,ik} \Sigma_{a\,k}{}^j )^{str} + {2 \over N-D}
\Sigma_a{}^{ij} \Theta^a
-  \left[ (M^2)^{ij} \right]^{str}= 0\,, &(5.11.c) \cr}
$$
where the symbol $(\cdots)^{str}$ denotes the
symmetric traceless part of the matrix under parenthesis,
and we have defined the `effective mass' matrix

$$
(M^2)^{ij} = - K_{ab}{}^i K^{ab}{}^j + g (R(E_a, n^i)E^a,n^j)
= (M^2)^{ji}\,. \eqno(5.12)
$$
This terminology is borrowed from the perturbative analysis of
Refs.[5,7], where it appears as a variable mass in the worldsheet
wave equation that describes the evolution of small perturbations.

The anti-symmetric Raychaudhuri equations, Eq. (5.4), give

$$
\eqalignno
{2 \tilde\nabla_{[a} \Lambda_{b]}{}^{ij} &= - 2 \Lambda_{[a}{}^{k[i}
 \Lambda_{b]}{}^{j]}{}_k - 2 \Sigma_{[a}{}^{k[i}
 \Sigma_{b]}{}^{j]}{}_k
- \Omega_{ab}{}^{ij}\,, &(5.13a) \cr
2 \partial_{[a} \Theta_{b]} &= 0\,, &(5.13b) \cr
2 \tilde\nabla_{[a} \Sigma_{b]}{}^{ij} &= - 2 ( \Lambda_{[a}{}^{ik}
\Lambda_{b]k}{}^j + \Sigma_{[a}{}^{ik}
\Sigma_{b]k}{}^j )^{str} + 4 \Lambda_{[a}{}^{k(i}
\Lambda_{b]k}{}^{j)}\,. &(5.13c)
\cr}
$$
First of all, note that if we set $\Lambda_a{}^{ij} ,
\Sigma_a{}^{ij}$ equal to zero initially, $\Lambda_a{}^{ij}$
will continue to vanish only if the curvature of
the extrinsic twist, $\Omega_{ab}{}^{ij}$, vanishes.
The generalized shear, $\Sigma_a{}^{ij}$, is picked up
if the matrix $(M^2)^{ij}$ has a non vanishing traceless part.

Let us focus now our attention on the equations which describe the
evolution of the generalized expansion, $\Theta_a$.
Eq. (5.13b) implies that
$$ \Theta_a = \partial_a \Upsilon\,, \eqno(5.14) $$
at least locally, for some potential function $\Upsilon$.
(However, recall that Eqs. (5.4), and thus Eqs. (5.13),
are vacuus for a curve.)
Inserting this expression in Eq. (5.11b), we find

$$
\Delta \Upsilon + {1 \over N-D} \partial_a \Upsilon \partial^a \Upsilon
- \Lambda^2 + \Sigma^2 - M^2 = 0\,, \eqno(5.15)
$$
where we have defined the worldsheet scalar quantities
$\Lambda^2 \equiv  \Lambda^{a\,ij} \Lambda_{a\,ij}$, $
\Sigma^2 \equiv  \Sigma^{a\,ij} \Sigma_{a\,ij} $, and $M^2 \equiv
(M^2)^i{}_i$. This equation describes the
evolution of the expansion of the worldsheet. It is
a quasi-linear, second order hyperbolic partial differential
equation. Note the non-linear term which depends quadratically
in worldsheet derivatives of $\Upsilon$. Neither this term
nor the other worldsheet scalars which follow it have a positive
sign, since the worldsheet metric has indefinite signature.

It is useful to compare Eq.(5.11b) for $\Theta_a$
(and its development through Eq.(5.14) to Eq.(5.15)) with
its one dimensional analogue. For a geodesic curve with tangent vector $V$,
the Raychaudhuri equation describing the evolution of the
expansion $\Theta$ is given by [8],
$$
\dot{\Theta} + {1 \over 3} \Theta^2 +
\Sigma^2 - \Lambda^2  + R_{\mu \nu} V^\mu V^\nu
= 0\,.\eqno(5.16)
$$
This is a first order ordinary differential equation. It can always
be converted into a second order equation by simply
redefining $\Theta =\dot\Upsilon$. Note, however, that
the non-linear term in $\Theta$ in Eq.(5.16)
is positive definite. In general relativity, when the
vorticity is set to zero, and the weak energy condition
holds,
$
R_{\mu \nu} V^\mu V^\nu \ge 0
$,
one comes to the conclusion that $\dot{\Theta}$
is negative, {\it i.e.} geodesics focus, and diverges within a
finite proper length [8]. It would be nice to be able to apply a
similar argument to Eq. (5.15), but it is not obvious to us
how to do this.

We can consider $\Upsilon$ as a generalized relative volume expansion
potential. If $l$ represents the characteristic length of the expansion,
we can set,

$$
\Upsilon = (N-D) \ln l \,, \eqno(5.17)
$$
With this elementary change of variables,
Eq. (5.15) translates into the linear equation,

$$
\Delta l + {1 \over N-D} \left[ \Sigma^2 - \Lambda^2 - M^2 \right] l
= 0\,. \eqno(5.18)
$$
This is a wave equation on the worldsheet for a massive {\it
positive definite} scalar field $l$ with an
effective mass term, $\mu^2 = - {1/N-D}[ \Sigma^2 - \Lambda^2 - M^2] $.
Superficially, we have reduced the analysis of $\Theta_a$ to the
solution of a linear wave equation. However, we need to remember that
$\mu^2$ involves $\Sigma_a^{ij}$ and
$\Lambda_a^{ij}$ explicitly as, as a result, depends implicitly also
on $\Theta_a$. We note that just as $M^2$ does not have a definite sign
neither does $\mu^2$.

Note that for a geodesic curve, with $\Theta = 3  (d / ds) (\ln l )$,
the geodesic Raychaudhuri equation Eq. (5.16) reduces to

$$
\ddot{l} + {1 \over 3} \left(
\Sigma^2 - \Lambda^2  + R_{\mu \nu} V^\mu V^\nu  \right)
l = 0\,. \eqno(5.19)
$$
In general relativity, using the same argument sketched above,
one finds that $ \ddot{l} $ is negative,
implying that $l$ cannot have a local minimum.

To conclude this section, we note that, for a hypersurface,
$ D = N - 1$, there is only one normal vector so that
$J_{a\,ij} = J_{a\,11} \equiv \Theta_a $.
The only degree of freedom describing the deformation of the hypersurface
is therefore the breathing mode or dilation of the
worldsheet. The extrinsic twist potential, $\omega_a{}^{11}$,
vanishes identically, by anti-symmetry, and the normal frame
rotation coefficients vanish as well. Is it also possible to
orient the tangent vectors along the normal direction so that
$S_{ab}{}^1 =0$, in a way analogous to a curve.

Therefore, the Raychaudhuri equations for this special case
reduce to Eq. (5.14), and

$$
\Delta \Upsilon + \partial_a \Upsilon \partial^a \Upsilon - M^2 = 0\,.
\eqno(5.20).
$$
The corresponding linear equation (5.18) is genuinely linear.

\vfill\eject
\noindent{\bf VI JACOBI EQUATIONS}
\vskip1pc

In this section, we show how the non-perturbative
generalized Raychaudhuri equations describing arbitrary
deformations of extremal membranes can be linearized to
reproduce the perturbative Jacobi equations describing small
deformations derived in Refs. [5,6,] and [7].

We demonstrate that, for an extremal membrane, the traced
Raychaudhuri equation, Eq.(5.3), alone completely encodes perturbation
theory. In fact, it is a simple matter to obtain the Jacobi equations
derived in Refs.[5,7]  directly from Eq.(5.3). The anti-symmetric
Raychaudhuri equations, Eqs. (5.4), turn out to have
a vacuous perturbative limit.

In Ref. [7], the infinitesimal deformations are decribed by a
multiplet of scalar fields, $\Phi^i$, living on the worldsheet.
The perturbations are characterized by the normal projections
of the infinitesimal perturbations of the worldsheet basis,

$$
J_a{}^i \equiv \Phi^j g ( D_j E_a , n^i)\,. \eqno(6.1)
$$
For infinitesimal perturbations, it was found that,

$$
J_a{}^i = \tilde\nabla_a \Phi^i\,. \eqno(6.2)
$$
Therefore, the reduction of the generalized Raychaudhuri equations to
their perturbative limit consists simply in the use of the relation,

$$
 J_a{}^{ji} \Phi_j = J_a{}^i = \tilde\nabla_a \Phi^i\,. \eqno(6.3)
$$

Let us consider now the (infinitesimal) normal  linear combination
of Eq.(5.3) obtained by contracting with the scalar fields $\Phi_j$,

$$
\left[ \tilde \nabla_a J^{a\,ji}
+ J_{a}{}^j{}_k J^{a\,ki} - (M^2 )^{ji} \right] \Phi_j = 0
\,,\eqno(6.4)$$
where we use the effective mass $(M^2 )^{ij}$ defined in Eq. (5.12).

This expression may be rewritten as,

$$
\tilde\nabla_a ( J^{a\,ji} \Phi_j ) - J^{a\,ji} (\tilde\nabla_a \Phi_j)
+ J_{a}{}^j{}_k J^{a\,ki} \Phi_j  - (M^2)^{ji} \Phi_j
= 0\,.
$$
Using Eq. (7.3), the second and the third term cancel, and one
recovers Eq. (5.3) of Ref. [7],

$$
\tilde\nabla^a \tilde\nabla_a \Phi^i - (M^2)^{i}{}_j \Phi^j = 0\,.
\eqno(6.5)$$

We show now that the anti-symmetric combination (5.4) is vacuus. To see this,
note that
with the use of the relation (6.3), one has,

$$
\left[ \tilde \nabla_b J_{a}{}^{ij} -\tilde \nabla_a J_{b}{}^{ij}
+ J_{b}{}^i{}_k J_{a}{}^{kj} - J_{a}{}^i{}_k J_{b}{}^{kj}\right] \Phi_i
=  \Omega_{ab}{}^{ij} \Phi_i\,. \eqno(6.6)
$$
However, the left hand side  of this equation gives

$$ \eqalign
{\tilde\nabla_b (J_a{}^{ij} \Phi_i) &- J_a{}^{ij} (\tilde\nabla_b \Phi_i)
- \tilde\nabla_a (J_a{}^{ij} \Phi_i) + J_a{}^{ij} (\tilde\nabla_b \Phi_i)
+ J_{b}{}^i{}_k J_{a}{}^{kj} \Phi_i - J_{a}{}^i{}_k J_{b}{}^{kj} \Phi_i \cr
&= 2 \tilde\nabla_{[b} \tilde\nabla_{a]} \Phi^j =  \Omega_{ba}{}^{ji} \Phi_i
=  \Omega_{ab}{}^{ij} \Phi_i\,, \cr }
$$
which is identical to the right hand side of  Eq. (6.6).

\vskip2pc
\noindent{\bf VII CONCLUSIONS}
\vskip1pc

We have provided a non-perturbative framework involving a coupled
system of non-linear partial differential equations to examine
the evolution of deformations of relativistic membranes of an
arbitrary dimension propagating in a background spacetime of
arbitrary co-dimension. The construction of this system of equations
was motivated by the Raychaudhuri equations describing point
particles to which they reduce when $D=1$. Despite the complexity of these
equations, they do share many features of the one-dimensional
prototype.  Clearly, however, work remains to be done to sharpen
our understanding of this system. Work is in
progress on the dynamics of large deformations
about various simple symmetrical configurations [11,13].

To conclude, we comment briefly on the derivation
of the appropriate generalizations of Raychaudhuri equations when
the dynamics is not of the Nambu type. Recall that to yield a useful
system of equations from the simple-minded generalization of
the Raychaudhuri equations Eq.(3.1), we need
to eliminate the source term $\tilde\nabla_i K_{ab}{}^j$, in terms of
quantities determined on the worldsheet by the equations
of motion. For the case of an extremal membrane, this was achieved
by forming appropriate linear combinations of Eq.(3.1), {\it i.e.}
by considering its trace over the worldsheet indices, Eq. (5.3).
We note, however, that the other linear combinations we considered,
the anti-symmetric combination Eq.(5.4), are in fact
independent of the dynamics, since the source term is
eliminated. Thus we continue to use these equations.
For a non-extremal surface, therefore, we need to find the
appropriate linear combination to replace Eq.(5.3).
The genealization of the procedure we followed for extremal membranes
is to consider linear combinations of Eq.(3.1)
such that any source term which is not determined on the
worldsheet is proportional to the equations of motion. If, for example,
rigidity corrections are incorporated into the action, the
dynamical equations will involve two worldsheet derivatives
($\tilde\nabla_a$) of $K_{ab}^i$ (four derivatives of the
embedding functions) [14]. This will require us to strike Eq.(3.1) twice
with $\tilde\nabla_a$. The source term will be removed in favor
of quantities which are determined completely on the worldsheet by
commuting these derivatives through the normal gradient operating
on $K_{ab}^i$. Details will be presented elsewhere [15].
%\vfill\eject
\vskip2pc
\centerline{\bf ACKNOWLEDGEMENTS}
\vskip1pc

Support from CONACyT under the grant 3354-E9308
is gratefully acknowledged.

\vskip2pc
\centerline{\bf APPENDIX A}
\vskip1pc

In this Appendix we give the details of the derivation
of the generalized Raychaudhuri equations  (3.1).
This involves the evaluation
of the quantity $\tilde\nabla_b J_{a\,ij}$.
We emphasize that we are not making  any
assumptions about the
equations of motion of the worldsheet.

The definition of $J_{a\,ij}$
in Eq.(2.13) implies
$$
\eqalign{D_b J_{a\, ij} =& D_b g(n_j, D_i E_a)\cr
			  =& g(D_b n_j, D_i E_a) +
			   g(n_j, D_b D_i E_a),.\cr}\eqno(A.1)$$
Using Eqs.(2.4) and (2.11), the first term gives us

$$
\eqalign{g(D_b n_j, D_i E_a )=&
g(K_{bc\,j} E^c + \omega_{b\,jk} n^{k},
S_{ad\,i} E^d + J_{a\,il} n^l)\cr
=& S_{ac\,i} K_b{}^c{}_j + \omega_{b\,j}{}^{k} J_{a\,ik}\,.\cr}\eqno(A.2)$$
We now apply the Ricci identity to the last term on the second line
of Eq.(A.1),

$$
D_b D_i E_a = D_i D_b E_a + R(E_b ,n_i)E_a +
(D_b n_i^\mu -D_i E^\mu{}_b )D_\mu E_a\,.\eqno(A.3)$$
We note that

$$
\eqalign{ D_i (D_b E_a )=&
D_i\left[\gamma_{ba}{}^c E_c  - K_{ba}{}^k n_k\right]\cr
=&  (D_i \gamma_{ba}{}^c) E_c + \gamma_{ba}{}^c D_i E_c
  - (D_i K_{ba}{}^k) n_k - K_{ba}^k D_i n_k\,.\cr}$$
Thus

$$
g(n_j, D_i (D_b E_a ))= + \gamma_{ba}{}^c J_{c\,ij}
- D_i K_{ab\,j} - \gamma_{ik\,j} K_{ab}{}^k \,.\eqno(A.4)$$
The first term appearing on the r.h.s. of Eq.(A.4) combines with the
LHS of Eq.(A.1) to provide a worldsheet covariant derivative of
$J_{a\, ij}$ defined by

$$
\nabla_b J_{a\,ij} = D_b J_{a\,ij} - \gamma_{ba}{}^c J_{c\,ij}\,.
\eqno(A.5)$$
The latter two terms add to yield a normal covariant derivative acting on
$K_{ab}{}^j$,

$$
\nabla_i K_{ba}{}^j =
D_i K_{ba}{}^j - \gamma_{i}{}^{jk} K_{ba\,k} \,.\eqno(A.6)$$
The term which remains to be evaluated is

$$
g(n_j, (D_b n^\mu{}_i - D_i E^\mu{}_b )D_\mu E_a)\,.$$
To evaluate it we insert unity to write

$$
\eqalign{g(n_j, (D_b n^\mu{}_i - D_i E^\mu{}_b )D_\mu E_a)=&
g(n_j, (D_b n_{\mu\,i} -D_i E_{\mu\,b})
(E^\mu{}_c E^{\nu\,c} + n^\mu{}_k n^{\nu\,k}) D_\nu E_a)\cr
=& g(n_j, (D_b n_{\mu\,i} - D_i E_{\mu\,b})
(E^{\mu\,c} D_c E_a + n^{\mu\,k} D_k E_a))\cr
=& \omega_{b\,i}{}^{k} J_{a\,kj} - K_{bc}{}^i K_a{}^{c\,j} +
S_{bc}{}^i K_a{}^{c\,j} - J_b{}^{ik} J_{a\,k}{}^j\,.\cr}\eqno(A.7)$$
Summing terms, we obtain the sought for expression in the form

$$
\tilde \nabla_b J_{a\,ij} =\tilde \nabla_i K_{ab\,j}
- J_{b\,i}{}^k J_{a\,kj}- K_{bc}{}^i K_a{}^{c\,j}
+  g (R ( E_b, n_i )E_a, n_j )
,,\eqno(A.8)$$
where we have exploited the definitions of $\tilde\nabla_b$
and $\tilde\nabla_i$, Eqs(2.9) and (2.17) respectively, to write

$$
\tilde \nabla_b J_{a\,ij} =\nabla_b J_{a\,ij}
- \omega_{b\,j}{}^{k} J_{a\,ki} - \omega_{b\,i}{}^{k} J_{a\,kj}
\,,\eqno(A.9)$$
and

$$
\tilde \nabla_i K_{ab}{}^j = \nabla_i K_{ab}^j -
K_{b }{}^{c\,j} S_{ac\,i} - K_{a}{}^{c\,j} S_{bc \,i}
\,.\eqno(A.10)$$

\vskip2pc
\centerline{\bf APPENDIX B}
\vskip1pc

In this appendix, we illustrate  how the integrability conditions
Eq. (5.7) imply that the spatial constraints (5.6.b) are preserved by the
evolution. We use a simpler analogous system of equations,
assuming that the worldsheet geometry is flat, and that
the extrinsic twist potential, $\omega_a{}^{ij}$, vanishes.
Under these assumptions, Eqs. (5.4) simplify to,
$$\partial_a J_b -\partial_b J_a = G_{ab}(J_c)\,.\eqno(B.1)$$
These equations fall into two categories. The first
consists of dynamical equations for the worldsheet spatial
components $J_A$,

$$
\partial_0 J_A -\partial_A J_0 = G_{0A}\,.\eqno(B.2)$$
The latter is a set of constraints on these variables,

$$
{\cal C}_{AB}\equiv \partial_A J_B -\partial_B J_A - G_{AB} =0 \,,
\eqno(B.3)$$
The integrability conditions for $(C.1)$ can be written as,

$$
\partial_c G_{ab}+\partial_a G_{bc} +
\partial_b G_{ca}=0\,.\eqno(B.4)$$
These equations are vacuous if two or more of the indices
are equal. To propagate the spatial constraint,
we require only the equation

$$
\partial_0 G_{AB} +\partial_A G_{B0}- \partial_B G_{A0} =0
\,.\eqno(B.5)$$
For now,

$$
\partial_0{\cal C}_{AB}=
\partial_A (\partial_0 J_B) -\partial_B
(\partial_0 J_A) - \partial_0 G_{AB}
\,,\eqno(B.6)$$
the right hand side of which is zero modulo Eqs.(B.2) and (B.4).
Thus ${\cal C}_{AB}=0$ is preserved by the evolution.

\vskip2pc
\centerline{\bf REFERENCES}
\vskip1pc

\item{1.} In cosmology, see
A. Vilenkin,  Phys. Rep. {\bf 121}, 263 (1985);
P. Shellard and A. Vilenkin {\it Cosmic Strings}  (Cambridge
Univ. Press, Cambridge, 1994)
\vskip1pc
\item{2.}
L. P. Eisenhart {\it Riemannian Geometry} (Princeton Univ. Press,
Princeton, 1947);
M. Spivak {\it Introduction to Differential Geometry} Vols. I and IV
(Publish or Perish, Boston MA 1970);
H. B. Lawson {\it Lectures on Minimal Submanifolds} 2nd. edition,
(Publish or Perish, 1980);
M. Dajczer, {\it Submanifolds and Isometric Immersions}
(Publish or Perish, Houston, Texas, 1990);
D.H. Hartley and R.W. Tucker,
in {\it Geometry of Low-Dimensional Manifolds: 1},  London
Mathematical Society Lecture Note Series {\bf 150},  ed. by
S.K. Donaldson and C.B. Thomas (Cambridge University Press, Cambridge,
1990); B. Carter, {\it Journal of Geometry and Physics}
{\bf 8}, 52 (1992).
\vskip1pc
\item{3.} For example, see  A. Polyakov, {\it Nucl. Phys.}
{\bf B268}, 406 (1986); T.L. Curtright, G.I. Ghandour, and C.K. Zachos,
{\it Phys. Rev.} {\bf D34}, 3811 (1986);
K. Maeda and N. Turok, {\it Phys Lett} {\bf B202} 376 (1988);
R. Gregory, D. Haws, and D. Garfinkle, {Phys. Rev.} {\bf D42}
343 (1990); R. Gregory, {\it ibid.} {\bf D43} 520 (1991);
B. Carter and R. Gregory, (Preprint, 1994).
\vskip1pc
\item{4.}
J. Garriga and A. Vilenkin,  Phys. Rev.  {\bf D44}, 1007 (1991);
{\it ibid.} {\bf D45}, 3469 (1992);{\it ibid.} {\bf D47}, 3265 (1993).
\vskip1pc
\item{5.} J. Guven, {\it Phys Rev} {\bf D48} 4606 (1993);
{\it ibid.} {\bf D48} 5562 (1993); in {\it Proceedings of
SILARG VIII}, ed. by P. Letelier and W. Rodriguez,
(World Scientific, Singapore, 1994).
\vskip1pc
\item{6.} B. Carter, {\it Phys. Rev.} {D48}, 4835 (1993).
A.L. Larsen and V.P. Frolov, {\it Nucl Phys} {\bf B414}, 129 (1994).
\vskip1pc
\item{7.} R. Capovilla and J. Guven, `` Geometry of
Deformations of Relativistic Membranes" preprint CIEA-GR-940,
ICN-UNAM-9405, gr-qc/9411060 {\it Phys. Rev.} {\bf D} (1995) (to appear).
\vskip1pc
\item{8.}
S.W. Hawking and G.F.R. Ellis {\it The Large Scale Structure of
Space-Time} (Cambridge Univ. Press, Cambridge, 1973).
\vskip1pc
\item{9.}
A.K. Raychaudhuri {\it Phys. Rev} {\bf 98} (1955) 1123
\vskip1pc
\item{10.} R. Capovilla and J. Guven, in the
proceedings of {\it The Seventh Marcel Grossmann Meeting on
General Relativity} ed. by R. Jensen (World Scientific, Singapore, 1995)
(to appear)
\vskip1pc
\item{11.} R. Capovilla and J. Guven (in preparation).
\vskip1pc
\item{12.} P.A.M. Dirac {\it Lectures on Quantum Mechanics}
(Academic Press, New York, 1964)
\vskip1pc
\item{13.} Progress in this direction has been reported by S. Kar,
`` Strings in a Wormhole Background" (gr-qc/9503004)
\vskip1pc
\item{14.} Recent work on spherically symmetric solutions include
P.S. Letelier, {\it Phys. Rev.}  {\bf D41} 1333 (1990);
B. Boisseau and P.S. Letelier, {\it ibid.} {\bf D46} 1721
(1992); C. Barrabes, B. Boisseau, and M. Sakellariadou, {\it ibid.}
{\bf D49} 2734 (1994).
\vskip1pc
\item{15.} R. Capovilla and J. Guven (in preparation).
\vskip1pc

\bye